\documentclass[pra,aps,twocolumn,10pt,showpacs,groupedaddress,superscriptaddress,floatfix]{revtex4-1}
\usepackage{amsmath,amsfonts,amssymb,graphics,graphicx,epsfig,color,times}
\usepackage{color}
\usepackage{mathrsfs}
\usepackage{verbatim}

\def\be{\begin{equation}}
\def\ee{\end{equation}}
\def\bea{\begin{eqnarray}}
\def\eea{\end{eqnarray}}
\def\ba{\begin{array}}
\def\ea{\end{array}}
\def\bdm{\begin{displaymath}}
\def\edm{\end{displaymath}}

\begin{document}
\title{Fragmented Many-body States of Spin-2 Bose Gas}
\author{H. H. Jen}
\affiliation{Institute of Physics, Academia Sinica, Taipei 11529, Taiwan}
\author{S.-K. Yip}
\affiliation{Institute of Physics, Academia Sinica, Taipei 11529, Taiwan}
\affiliation{Institute of Atomic and Molecular Sciences, Academia Sinica, Taipei 10617, Taiwan}

\date{\today}

\renewcommand{\r}{\mathbf{r}}
\newcommand{\f}{\mathbf{f}}

\def\be{\begin{align}}
\def\ee{\end{align}}
\def\bea{\begin{eqnarray}}
\def\eea{\end{eqnarray}}
\def\ba{\begin{array}}
\def\ea{\end{array}}
\def\bdm{\begin{displaymath}}
\def\edm{\end{displaymath}}
\def\red{\color{red}}

\begin{abstract}
We investigate the fragmented many-body ground states of a spin-2 Bose gas in zero magnetic field.\ We point out that the exact ground state is not simply an average over rotationally-invariant mean-field states, in contrast to the spin-1 case with even number of particles N.\ We construct the exact ground states and compare them with the angular-averaged polar and cyclic states.\ The angular-averaged polar states fail to retrieve the exact eigenstate at $N$ $\ge$ $6$ while angular-averaged cyclic states sustain only for N with a multiple of $3$.\ We calculate the density matrices and two-particle density matrices to show how deviant the angular-averaged state is from the exact one.
\end{abstract}
\maketitle
{\it Introduction.--} Since the advancement of optically trapped Bose-Einstein condensate (BEC) \cite{Stamper-Kurn1998}, spinor BEC \cite{Stenger1998} has provided a paradigm to study magnetism, spin textures, topological excitations, and quantum dynamics of associated many-body ground states \cite{Kawaguchi2012, Stamper-Kurn2013}.\ The quantum phases of spin-$f$ BEC can be ferromagnetic or polar for $f$ $=$ $1,2$ \cite{Ho1998, Ohmi1998}, or cyclic ({\it f} $=$ $2$)
\cite{Ciobanu2000,Koashi2000,Ueda2002} depending on two-body $s$-wave scattering lengths $a_F$ of even total spin $F'$ up to $2f$.\
Uniaxial and biaxial spin nematic phases can also be identified in spin-2 case when the degeneracy in polar phase is lifted by thermal or quantum fluctuations \cite{Turner2007,Song2007}.
Higher spin Bose gas can involve even more complicated phases \cite{Diener2006, Santos2006, Yip2007, Kawaguchi2011}.
 \ Rich spin mixing dynamics has been used to observe the ferromagnetic \cite{Chang2004} and anti-ferromagnetic (polar) properties \cite{Black2007} respectively for spin-1 $^{87}$Rb and $^{23}$Na Bose gases.\ In spin-2 cases, polar phase is the likely phase for $^{87}$Rb \cite{Schmaljohann2004, Kuwamoto2004}.\ Recent experimental developments
in spinor BEC involve spin textures \cite{Guzman2011}, spin dynamics \cite{Kronjager2006, Klempt2009}
under quadratic Zeeman shift.
 There are also studies of quantum phase transitions by Faraday rotation spectroscopy \cite{Liu2009} or adiabatic microwave fields \cite{Jiang2014}, and spin coherence measurements by Ramsey interferometry \cite{Sadgrove2013, Eto2014}.

A condensate of bosons forms when one of its single-particle wavefunction is macroscopically occupied \cite{Penrose1956}.\ The fragmentation of BEC becomes feasible when multiple macroscopic single-particle densities are degenerate in spinor Bose gases \cite{Nozieres1982} though it is fragile in the presence of weak external magnetic fields or symmetry-breaking perturbations \cite{HoYip2000}.\
Lately many interests in fragmented BEC include dynamical formation of two-dimensional fragmented BEC \cite{Klaiman2014}, quadratic Zeeman effect on spin fragmentation \cite{Tasaki2013, Sarlo2013}, and fragmented many-body ground states with anisotropic long-range interactions \cite{Bader2009} or trapping potentials \cite{Cizek2013}.\
It has been proposed that signatures of fragmentation can be probed by measuring density-density correlations \cite{Kang2014} while fragmentation resulting from Goldstone magnon instability \cite{Kawaguchi2014} and spin-orbit coupling \cite{Ozawa2012-Zhou2013-Song2014} are also investigated.

The fragmented structure of the ground state in spin-1 Bose gases \cite{HoYip2000, Law1998, Mueller2006} originates from the rotational invariance in spin degrees of freedom.
 The symmetry-breaking mean-field (MF) treatment fails in describing the exact ground state for presumption of single spin coherent condensate.
  For scattering lengths obeying $a_2>a_0$, the MF state is polar, but the exact ground state is fragmented and, for even number of particles N,
  can be viewed as a collection of two-particle spin-singlets.
This exact ground state  has equal populations in the magnetic sublevels with large number fluctuations of order N \cite{HoYip2000}, which is very different from MF states.\ It is claimed \cite{Mueller2006} that this exact ground state can be understood as the angular-average of the MF polar states as an analog to the relation between Fock and coherent states in a double-well system \cite{Mueller2006}.\ This remains the view adopted by the most recent review articles
  \cite{Kawaguchi2012, Stamper-Kurn2013}. Is this perspective of angular-averaged states universal and applicable in constructing the exact ground states for larger spins?\ In this paper, we investigate many-body ground states of a spin-2 Bose gas, and demonstrate how angular-averaged states are unable to construct them.\ That the angular averaged MF states is the exact ground state is just a coincidence in spin-1 system.\ We address the inapplicability of the angular-averaging process, and also show how the angular-averaged MF state deviates from the exact eigenstates by studying the two-particle density matrices.

\begin{figure*}[t]
\centering
\includegraphics[width=16cm,height=8cm]{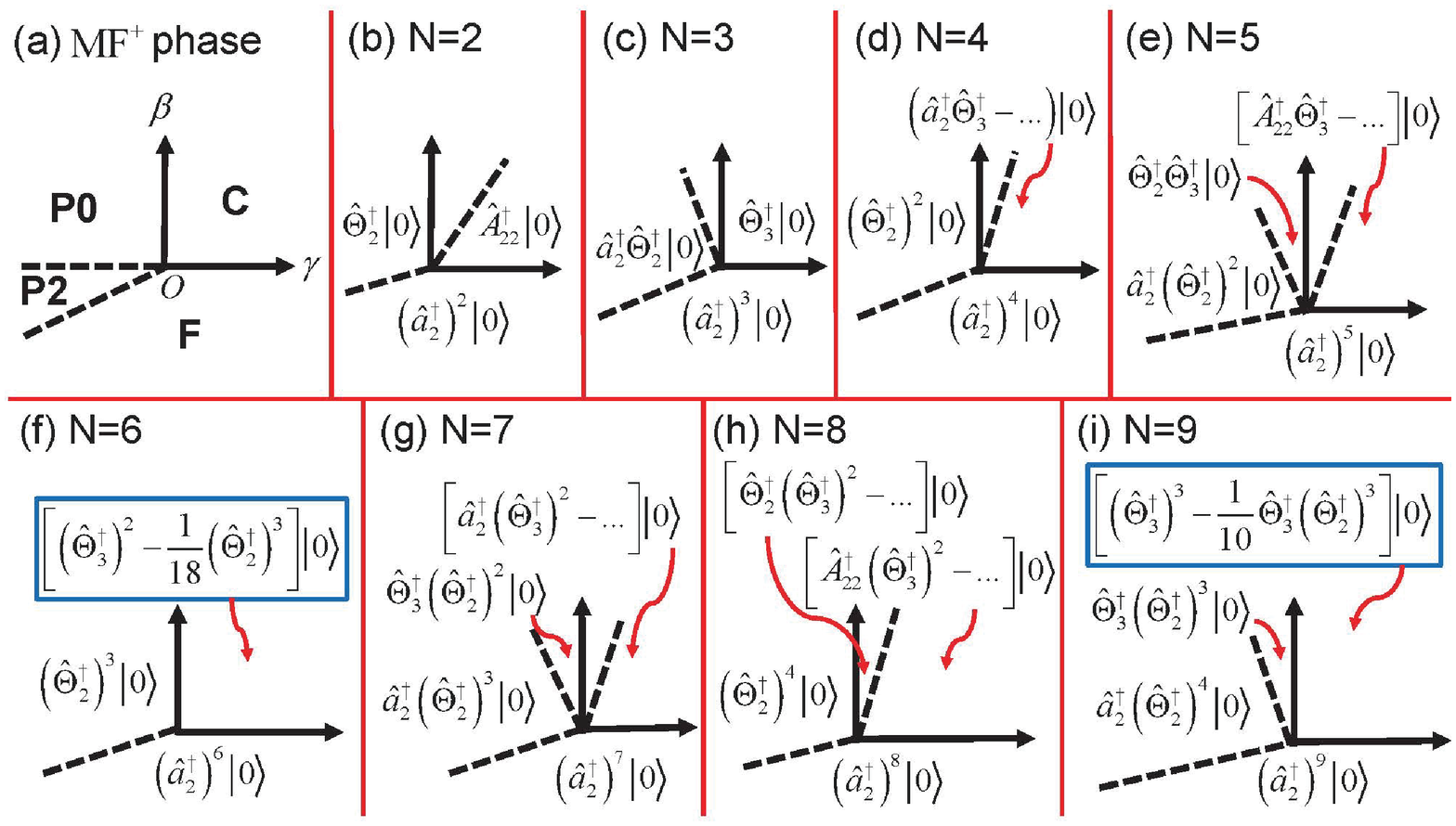}
\caption{(Color online) Mean-field-plus (MF$^+$) phase diagram and many-body ground states of spin-2 Bose gas in parameter space $(\gamma,\beta)$.\ (a) Mean-field-plus phase diagram: ferromagnetic (F), polar (P0 and P2), and cyclic (C) phases.\ (b-i): phase diagram for exact many-body states with finite number of particles from $N$ $=$ $2-9$. The form of the many-body ground states are shown. These states are constructed by spin-singlet pairs ($\hat{\Theta}_2^\dag$) and trios ($\hat{\Theta}_3^\dag$),
 and in certain regions, creation operator $a_2^\dag$ ($A_{22}^\dag$) for a single particle (pair of particles) with $f=2, m=2$ are also necessary (see text for definition of these operators).
  States that involve $\hat a_2^\dag$ or $\hat A_{22}^\dag$ explicitly are degenerate with their
 partners obtained by rotational symmetry (not shown). Dots in the formulas, when the complete expressions are not given, indicate that linear superposition with other terms are required.  Wavefunctions shown here are not normalized.
 \ Dashed lines indicate the (schematic) phase boundaries.}
 \label{fig1}
\end{figure*}

{\it Spin-2 Bose gas.--} For a spin-$f$ Bose gas at low temperature, the two-body particle interaction involves only scattering channels of even total hyperfine spin $F'$ states up to $2f$ \cite{Ho1998}.
We shall consider the single-mode approximation (SMA) where
 the spatial part of the wavefunction is the same for all spin sublevels such that the field operator $\hat{\psi}_m(\r)$ $=$ $\sqrt{\rho(\r)}\hat{a}_m$ with the density $\rho(\r)$ and spinor operator $\hat{a}_m$.\
 Since the spatial part is frozen, the effective Hamiltonian (in zero magnetic field,
 to which we shall limit ourselves) involves only the interaction $V$ which reads  \cite{Ciobanu2000}
\begin{equation}
\begin{split}
V&=\frac{1}{2}\int d\r \rho^2\Big(\sum_{m,m'=-2}^2\alpha \hat{a}^\dagger_m \hat{a}_{m'}^\dagger \hat{a}_{m'} \hat{a}_m \\
&+\sum_{\substack{m,n,m',\\n'=-2}}^2 \beta\hat{a}_m^\dagger \hat{a}_{m'}^\dagger\f_{mn}\cdot\f_{m'n'}\hat{a}_{n'}\hat{a}_n\\
&+\sum_{\substack{m,n,m',\\n'=-2}}^2 5\gamma\hat{a}_m^\dagger\hat{a}_{m'}^\dagger\langle 2m;2m'|00\rangle\langle00|2n;2n'\rangle\hat{a}_n\hat{a}_{n'}\Big),\label{V}
\end{split}
\end{equation}
where the coefficients are $\alpha$ $=$ $(4g_2+3g_4)/7$, $\beta$ $=$ $(g_4-g_2)/7$, and $\gamma$ $=$ $(g_0$ $-$ $g_4)/5$ $-$ $2(g_2-g_4)/7$.
Here the interaction parameters $g_F$ $\equiv$ $4\pi\hbar^2 a_F/M$ with the mass of the atom $M$ and
s-wave scattering length $a_F$,
and $\langle00|2n;2n'\rangle$ is the Clebsh-Gordan coefficient for the overlap between the states with two spin-2 bosons of $m_z$ $=$ $n,~n'$ and the spin singlet $|00\rangle$.

In MF theory, bosons condense.  Particles macroscopically occupy a single quantum state
which can be described by a spin-$2$ wavefunction $(\varphi_{-2}, ..., \varphi_{2})$.   In our case, there are
three phases characterized by two order parameters of magnetization $\langle\hat{f}\rangle$ $\equiv$ $\sum_{m=-2}^{2}m\varphi_m^*\varphi_m$ and spin-singlet pair amplitude $\langle\hat{\Theta}_2\rangle$ \cite{Ciobanu2000, Koashi2000}, where
 \begin{align}
 \hat{\Theta}_2 & \equiv\sum_{m=-2}^2 \sqrt{5} \langle 00 |2 m;2~-m \rangle \hat a_m \hat a_{-m} \nonumber \\
 &=  2\hat{a}_2\hat{a}_{-2} - 2\hat{a}_1\hat{a}_{-1} + \hat{a}_0^2
  \label{T2}
  \end{align}
is an operator which annihilates a singlet pair.
That is, we have $ \langle\hat{\Theta}_2\rangle$ $=$ $\sum_{m=-2}^{2} (-1)^m \varphi_m \varphi_{-m}$.
There are three phases.
The ferromagnetic (F) phase has a finite $\langle\hat{f}\rangle$ and zero $\langle\hat{\Theta}_2\rangle$, while the polar phase (P) has $\langle\hat{\Theta}_2\rangle = 1$ without $\langle\hat{f}\rangle$.\ When $\beta$, $\gamma$ $>0$, the cyclic (C) phase has the lowest mean-field energy for both zero $\langle\hat{\Theta}_2\rangle$ and $\langle\hat{f}\rangle$, breaking the time-reversal symmetry.\ The phase boundary between F and polar phases is delineated by the line  $4\beta$ $=$ $\gamma$.
\ The phase diagram is as shown in Fig. \ref{fig1}(a) \cite{Ciobanu2000, Koashi2000}.  Representative wavefunctions are
$(1,0,0,0,0)$ for F; $(1,0,0,\sqrt{2},1)$ which is equivalent by rotation to $(1,0,\sqrt{2},0,-1)$ for C \cite{Yip2007}.
Within mean-field, the polar phase P can have wavefunctions P0 $=$ $(0,0,1,0,0)$ or P2 $=$ $(1,0,0,0,1)$,
or any real linear combinations thereof \cite{Mermin1974} (apart from rotations).
This degeneracy however is lifted by fluctuations \cite{Turner2007, Song2007}.  The resulting phase diagram, which we shall call the  mean-field-plus (MF$^+$) phase diagram, is
shown in Fig \ref{fig1}(a). \ We note that P0(2) can be also represented by the polynomial forms via spherical harmonics as $(2z^2-x^2-y^2)$ and $(x^2-y^2)$ respectively \cite{Mermin1974,Yip2007}.\
For ease of referral later, we shall call the regions in $(\beta, \gamma)$ parameter space
occupied by the F, C, P phases as F, C, P  regions.

{\it Many-body ground states.--} Let us now discuss the many-body ground states of a spin-2 Bose gas in zero magnetic field with SMA. \cite{HoYin2000,Koashi2000, Uchino2008}.\ For a given N, the many-body ground states are characterized by two quantum numbers $F$ and $\tau$.\ $F$ is the total spin and the integer quantum number $\tau$ can be interpreted as the number of particles other than spin-singlet pairs, therefore $\tau$ is given by $3n_{30}$ $+$ $\lambda$ \cite{Koashi2000, Uchino2008} where $n_{30}$ is the number of spin-singlet trios and the integer $\lambda$ indicates particles other than spin-singlet pairs and trios.\ These states are also eigenstates of the operator $\hat \Lambda \equiv \hat \Theta_2^\dag \hat{\Theta}_2$ with eigenvalues $\Lambda = N(N+3)-\tau(\tau+3)$ $=$ $(N-\tau)(N+3 + \tau)$, from which $\tau$ can be evaluated.\ The exact ground state energy is proportional to $\beta F(F+1)$ $-$ $\gamma\tau(\tau+3)$ aside from a term depending only on $N$.\ The phase diagram derived by minimizing the ground state energy is as sketched in Fig. \ref{fig1} for $N$ $=$ $2$ to $9$.\ The line that separates the two phases in the $\beta$ $<$ $0$, $\gamma$ $<$ $0$ region is given by $(4N+2)\beta/(N+3)$ $=$ $\gamma$ \cite{Uchino2008}
  which approaches to MF phase boundary in thermodynamic limit.

The wavefunctions listed in Fig \ref{fig1} are constructed
according to the eigenvalues $F$ and $\tau$ which minimize the energy.
In the ferromagnetic region F, they are $(\hat a_2^{\dag})^N|0 \rangle$, hence
identical with the mean-field states.  (Here $|0 \rangle$ denotes the vacuum).
 For the polar P region, wavefunctions differ
according to whether $N$ is even or odd. For even $N$, the wavefunctions are $ (\hat \Theta_2^\dag)^{N/2} | 0 \rangle$
corresponding to $\tau=0$, $F$ $=$ $0$, and maximum possible $\Lambda$'s which are $N(N+3)$ (see also \cite{SM}).
For odd $N$,  the states that appear near the $- \gamma$ axis have wavefunctions
$ \hat a^\dag_2 (\hat \Theta_2^\dag)^{(N-1)/2} | 0 \rangle$ (up to rotations) again correspond to states with
maximum possible $\Lambda$'s which are now $ (N-1)(N+4)$ with $\tau$ $=$ $1$ and $F$ $=$ $2$.
Near the $+\beta$ axis but still $\gamma < 0$, the states (for $N \ge 3$) have the form
$\hat \Theta_3^\dag (\hat{\Theta}_2^\dagger)^{(N-3)/2}|0\rangle$ where
\begin{align}
\hat{\Theta}^{\dag}_3 &=  -\sqrt{6} \hat a^\dag_2 \hat a^\dag_0 \hat a^\dag_{-2}
 +\frac{3}{2} (\hat a^\dag_1)^2  \hat a^\dag_{-2} + \frac{3}{2} (\hat a^\dag_{-1})^2  \hat a^\dag_{2}  \nonumber \\
 &  - \sqrt{\frac{3}{2}} \hat a^\dag_1 \hat a^\dag_0 \hat a^\dag_{-1}
 +\frac{1}{\sqrt{6}} (\hat a^\dag_0)^3
\label{T3}
\end{align}
is an operator which creates a spin-0 trio (not normalized). These states have $\tau$ $=$ $3$
(see the end of Sec IV in \cite{SM}) and
$\Lambda$ $=$ $(N-3)(N+6)$ with $F$ $=$ $0$.

In the region C, if the particle numbers are multiple of $3$, the ground state wavefunctions are singlets
constructed by $\hat \Theta^\dag_3$ and $\hat \Theta^\dag_2$ with the eigenvalue $\Lambda$ $=$ $0$.
The wavefunctions are indicated in Fig \ref{fig1} and their derivation can be found in Supplemental Material \cite{SM}.
For other $N$'s, the C region is divided into two parts.  The states near the $+\beta$ axis and $\gamma > 0$
are again spin singlets. If $N$ $=$ $2 ({\rm mod} 3)$, hence $N$ $=$ $3R + 2$, they are given by $\hat \Theta^\dag_2 |\Psi_{3R} \rangle$ (with $\tau$ $=$ $3R$ and $F$ $=$ $0$) where $|\Psi_{3R} \rangle$ is the corresponding ground state in region C for $3R$ particles.
For $N$ $=$ $1 ({\rm mod} 3)$ and hence $N$ $=$ $3R + 4$ for integer $R$ (when $N \ge 4$)
 then the wavefunctions are $ (\hat \Theta^\dag_2)^2 |\Psi_{3R} \rangle$ and again $\tau$ $=$ $3R$.
 For the $N$'s shown in Fig \ref{fig1}, these states happen to be the same as the states near the $+\beta$ axis
 on the $\gamma < 0$ side so that they are the same phase, but this needs not hold
 for larger particle numbers $N \ge 11$.
 The states near the $+\gamma$ axis with $\beta > 0 $ have instead finite magnetization
 and are not rotationally invariant.  They are not directly relevant in the rest of
 the paper and we shall not discuss them in detail.

{\it Angular-averaged mean-field states.--}
Now we turn to the angular-averaged MF states and compare with
the corresponding exact eigenstates.
First we recall  the corresponding results for spin-1.  The MF ferromagnetic state
corresponds to the exact solution.  Their angular average actually vanishes.
 The polar mean-field state has a finite average only for $N$ even, and give the
 correct exact many-body state \cite{Mueller2006}.
 We then demonstrate how for spin-2 the averaging process enables the fragmentation in both the polar and cyclic phases but in general it fails to correctly construct the corresponding exact eigenstates at a given point $(\beta,\gamma)$
 in parameter space.

The situation for the ferromagnetic state is exactly the same as the spin-1 case.
We now consider the angular-averaged polar state of P0(2).  Starting from
the reference state $(0,0,1,0,0)$, the state obtained by rotations via
the Euler angels $\alpha,\beta,\gamma$, which we denote collectively as $\hat \Omega$, is given by
$\varphi_m^{P0} (\hat \Omega) = D^{(2)}_{m,0}  (\hat \Omega)$ where
the matrix $D^{(2)}_{m, m'}(\hat \Omega)$ is the spin-2 irreducible representation of the rotation operator \cite{Brink1968}
 (see also Supplemental Materials \cite{SM}).
 The general (unnormalized) rotationally invariant state is constructed via
\begin{align}
\left|\Psi \right\rangle_{av}=  \frac{1}{\sqrt{N!}} \int_{\hat \Omega}
(\hat a^\dag (\hat \Omega))^N   \ | 0 \rangle
\label{angular}
\end{align}
where we have defined $\int_{\hat \Omega} \equiv \int_0^{2 \pi} \frac{d \alpha}{ 2 \pi}
\int_0^\beta \frac{ d \beta \sin \beta}{ \pi} \int_0^{2 \pi} \frac{ d \gamma} { 2 \pi}$.
For our polar state P0, we thus use
$\hat a^\dag (\hat \Omega) \to
\sum_m \hat a^\dag_m \varphi_m^{P0} (\hat \Omega)$.  We call the resulting state $| \Psi_{P0} \rangle_{av}$.
It is straight-forward to evaluate the angular integrals.
 We find that  $| \Psi_{P0} \rangle_{av}$ retrieves the exact ground states for even $N$ $=$ $2,4$.
  For odd $N$ $=$ $3,5,7$, we recover the spin singlet state located near the $+\beta$ axis
   as indicated in Fig. \ref{fig1}
(see more details in \cite{SM}).
(obviously the angular average cannot produce the states with finite magnetization near the $- \gamma$ axis).
\ However for even $N$ $\geq$ $6$ and odd $N$ $\geq$ $9$, the angular-averaged states fail to construct the exact ground states.\  For example, for $N=6$ it gives rather
\begin{align}
\left|\Psi_{P0}\right\rangle_{av}&=\frac{1}{7 \cdot 11 \cdot 13 \sqrt{6!}}
\bigg[ 5 \cdot 3^2 \left(\hat{\Theta}_2^\dagger\right)^3\nonumber\\&
+ 3 \cdot 2^4\left(\hat{\Theta}_3^\dagger\right)^2\bigg]|0\rangle \ , \nonumber
\end{align}
which is in fact not even an eigenstate of the Hamiltonian in Eq. (\ref{V}).

Similarly for P2, we can construct angular-averaged states as in Eq. (\ref{angular})
except now we use $\hat a^\dag(\hat \Omega) \to
 \sum_m \hat a^\dag_m \varphi_m^{P2} (\hat \Omega)$
with $\varphi_m^{P2} (\hat \Omega) = \frac{1}{\sqrt{2}} (D^{(2)}_{m 2} + D^{(2)}_{m,-2})  (\hat \Omega)$.
In this case the angular average vanishes if $N$ is odd.  For $N$ even again it
produces the correct ground states for $N=2,4$ but fails again at $6$.

Actually why the angular averaged mean-field states can or cannot produce the
manybody state is now clear.  For $N$ up to $5$, the exact many-body singlet states are unique.
Since angular averaged mean-field states must either be zero or they must
be a rotationally invariant, they must either vanish or produce the singlet states.
This is actually independent of whether the starting mean-field state is
the corresponding ground state for the given parameters in the Hamiltonian.
For $N=6, 8, 9 ..$, the many-body singlet states are no longer unique.  The angular
average, if it is not zero, just produces some linear combinations of
these singlets.  The resulting states have nothing to do with the ground state
solutions of the Hamiltonian.   That the angular average of
the polar state for spin-1 produces correctly the exact many-body state
for even $N$ is purely because that, for spin-1, this singlet is unique.

Let us also consider the angular-averaged states for a linear combination of both P0(2), and use
$\varphi_m^P(\hat{\Omega})$ $\equiv$ $ {\rm cos} \theta D^{(2)}_{m,0}(\hat{\Omega})
+ \sin \theta (D^{(2)}_{m,2}(\hat{\Omega})+D^{(2)}_{m,-2}(\hat{\Omega}))/\sqrt{2}$ in Eq. (\ref{angular}).
For $N$ $=$ $6$, and the angular-averaged polar state becomes (see \cite{SM})
\begin{align}
\left|\Psi_P(\theta)\right\rangle_{av}&=\frac{1}{1001\sqrt{6!}}\bigg[(47-2\cos 6\theta)\left(\hat{\Theta}_2^\dagger\right)^3\nonumber\\&
+(12+36\cos 6\theta)\left(\hat{\Theta}_3^\dagger\right)^2\bigg]|0\rangle,\label{polar_6}
\end{align}
which in general again fails to become an eigenstate of Eq. (\ref{V}).
 The $\theta$ dependence of the coefficients obtained above can be understood by the symmetries of the general polar state under $\theta\to-\theta$ and $\theta\to\pi/3-\theta$ \cite{SM}.\ We note that $\left|\Psi_P(\theta)\right\rangle_{av}$ never produces the many-body state $[(\hat \Theta^\dag_3)^2 - \frac{1}{18} (\hat \Theta^\dag_2)^3 ]| 0 \rangle$ in the C region.\ It happens that when $\cos 6\theta$ $=$ $-1/3$, the angular-averaged polar state becomes the exact ground state $(\hat \Theta^\dag_2)^3 | 0 \rangle$ in the P region.\

The above special value of $\theta$ can be understood as follows.
It can be shown that the weighted average $3/2\int_0^{\pi/3}d\theta\sin3\theta$ over $\theta$,
 together with the average over Euler angles above, is equivalent to an average over the 4-sphere
  in the quantum rotor picture of \cite{Barnett2011}.    If we apply this average to
  $\varphi_m^P (\hat \Omega)$, we obtain the exact many-body state $ (\Theta_2^\dag)^{N/2} | 0 \rangle$
  for even $N$ (the average vanishes for odd $N$). \
  This is because the above mentioned averages guarantee that we obtain a state that is
  invariant under SO(5) rotations, and $ (\Theta_2^\dag)^{N/2} | 0 \rangle$ is the only such
  state (corresponding to $\tau= 0$ of  \cite{Uchino2008}).
   The $\theta$-averaged of $\rm cos( 6 \theta)$ is $-1/3$.

The comparison between the angular-averaged polar state and the exact eigenstates can also be viewed in a different manner.\ Let us consider the operator  $\hat \Theta_2^{\dag} \hat \Theta_2$.\ We note that, for the polar state, $\sum_m (-1)^m \varphi_m(\hat \Omega) \varphi_{-m}(\hat \Omega)$ $=$ $1$ for any $\hat \Omega$ and hence, for the normalized state $| \tilde \Psi_p (\theta) \rangle_{av} $ of $|\Psi_p(\theta)\rangle_{av}$, we have the expectation value
\begin{align}
_{av}\langle \tilde \Psi_p (\theta) |\hat \Theta_2^{\dag} \hat \Theta_2 |\tilde \Psi_p (\theta)\rangle_{av} = N (N-1) X_{N-2}(\theta) / X_N(\theta),\label{EN}
\end{align}
where $X_N(\theta)$ $\equiv$ $_{av}\langle\Psi_p(\theta)|\Psi_p(\theta)\rangle_{av}$ and we have defined $X_0 =1$.\ While this formula is general, let us focus on $P0$.\ The evaluation of $X_N$ at $\theta$ $=$ $0$ are particularly straightforward.\ We have
\begin{align}
X_N(0)\equiv~_{av}\langle\Psi_p(0)|\Psi_p(0)\rangle_{av} &=\int_{\hat \Omega} d \hat \Omega \left[D^{(2)}_{0,0}(\hat \Omega)\right]^N,
\end{align}
where $\hat \Omega$ $\equiv$ $\hat \Omega_1^{-1}\hat \Omega_2$
represents the rotation $\hat \Omega_2$ followed by the inverse of
$\hat \Omega_1$. Here we have used the relation $D^{(2)}_{0,0}(\hat \Omega)$ $\equiv$
$\sum_m$ $D_{m,0}^{(2)*}(\hat \Omega_1)D_{m,0}^{(2)}(\hat \Omega_2)$
\cite{Brink1968}.\
We obtain $X_1= 0$, $X_2 = 1/5$, $X_3 = 2/35$, $X_4 = 3/35$, $X_5 = 4/77$, $X_6 = 53/(7 \cdot 11 \cdot 13)$,
 $X_7 = 6/(11 \cdot 13)$, $X_8 = 5 \cdot 19/ ( 11 \cdot 13 \cdot 17)$, $X_9 = 2^3 \cdot 197 / ( 11 \cdot 13 \cdot 17 \cdot 19)$.
On the other hand,
as already mentioned, the exact eigenstates are also eigenvectors of the operator $\hat \Lambda \equiv \hat{\Theta}_2^\dag \hat{\Theta}_2$.\ For even N, the states $ (\hat{\Theta}_2^\dag)^{N/2} | 0 \rangle $  have eigenvalues $\Lambda$ $=$ $N(N+3)$.\ For odd N, the exact eigenstates $ (\hat{\Theta}_2^\dag)^{(N-3)/2} \hat{\Theta}_3^\dag |0 \rangle $  have eigenvalues
$(N-3)(N+6)$.\ We can check directly from Eq. (\ref{EN}) that the expectation values for $\hat{\Theta}_2^{\dag}\hat{\Theta}_2$ equal these exact values for $N$ $=$ $2$, $3$, $4$, $5$, $7$ but not $6$, $8$ or $9$.\ 

{\it Cyclic states.--} We now study the angular-averaged cyclic states. \ It is
simplest to use the reference state $  \frac{1}{\sqrt{3}} (1,$ $0,$ $0,$ $\sqrt{2}, 0)$ and hence
 $\varphi_m^C (\hat \Omega)$ $=$  $\frac{1}{\sqrt{3}}(D^{(2)}_{m, 2}$ $+$ $\sqrt{2} D^{(2)}_{m,-1})(\hat \Omega)$ in Eq. (\ref{angular}).
The angular-averaged C states are finite only when $N$ is a multiple of $3$, which can be easily seen
by considering the integral over the angle $\gamma$.
 It turns out that, in these cases, the angular-averaged C states do produce the correct
 many-body states.  This is due to the fact that $\varphi_m^C (\hat \Omega)$ obeys
 $\sum_m (-1)^m \varphi_m (\hat \Omega) \varphi_{-m} (\hat \Omega) = 0$, and hence
 the angular averaged state is annihilated by $\hat \Theta_2$, so that the resulting
 state must satisfy $\Lambda =0$ and hence correctly produce the corresponding many-body state.
 We have also verified this conclusion by direct angular averages (see \cite{SM}).
\begin{figure}[t]
\includegraphics[width=8.5cm,height=5cm]{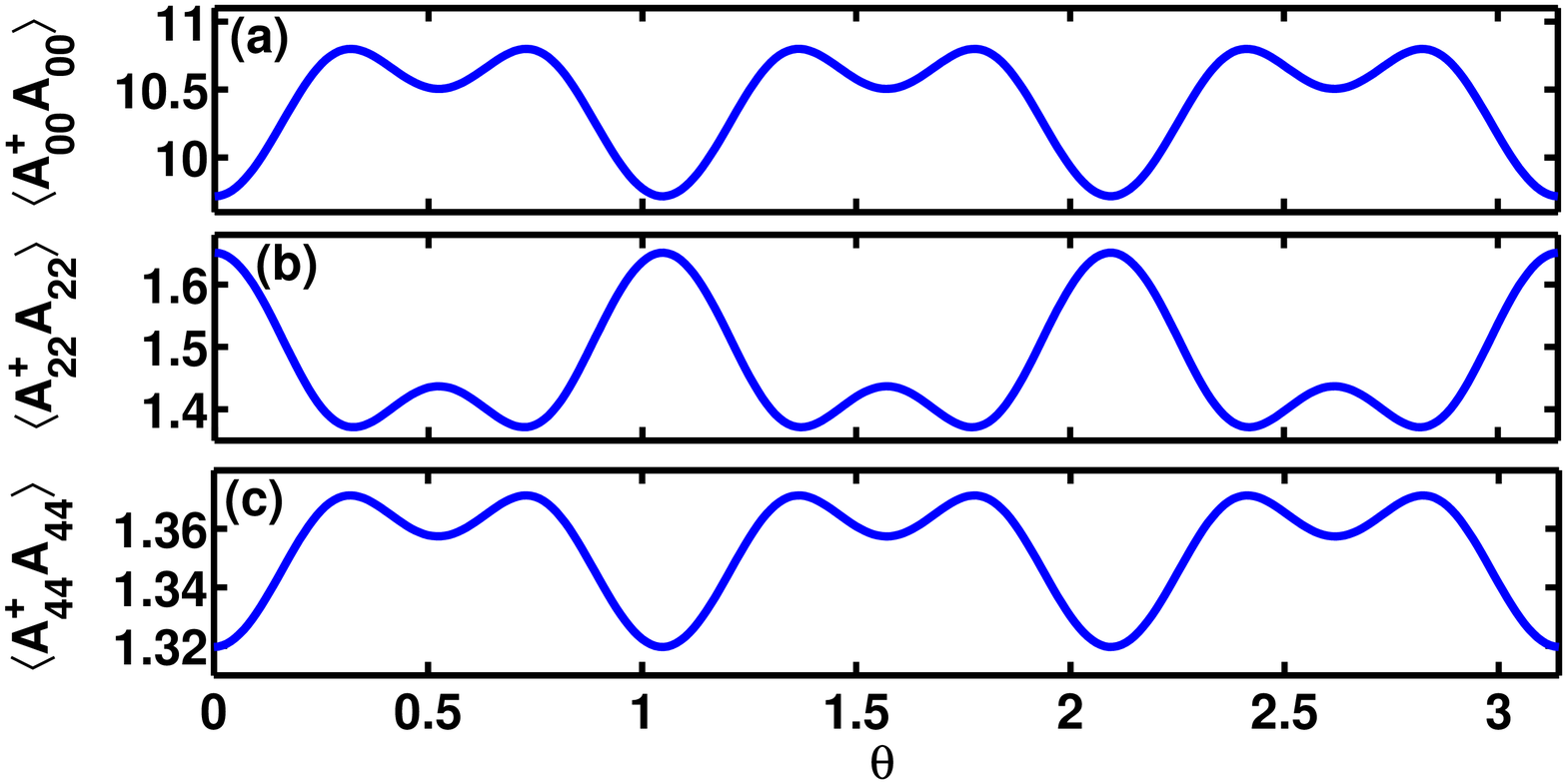}
\caption{(Color online) Two-particle density matrices for the angular-averaged polar state for $N$ $=$ $6$.\ Three density matrices for $J$ $=$ $0$, $2$, $4$ are shown in (a), (b), and (c) respectively.
}\label{fig2}
\end{figure}

{\it Reduced density matrix.--} As a further investigation,
compare, for $N=6$, the two-particle density matrices for the angular-averaged polar states
 with those for the exact many-body states
$|\Psi_{6} \rangle \equiv \frac{1}{\sqrt{g(3)}}
(\Theta^\dag_2)^3 | 0 \rangle $
where $g(3)$ is a normalization constant.
(The one-particle density matrices are obviously
identical since both states are rotational invariant).
   It is simplest to
     present the results using the operators
     \begin{align}
     \hat A_{JM} \equiv \sum_{m_1,m_2} \langle JM | 2 2 m_1 m_2 \rangle \hat a_{m_1} \hat a_{m_2} \ .
     \end{align}
      $ \langle \Psi_{6} | \hat A_{JM}^\dag \hat A_{J'M'} | \Psi_{6} \rangle $
      is finite only when $J=J'$and $M=M'$, and is further $M$ independent, as expected by rotational
      invariance.   These values are discussed in \cite{SM}.
          We have
      $\langle \Psi_{6} | \hat A_{00}^\dag \hat A_{00} | \Psi_{6} \rangle
         = 54 / 5  = 10.8$, and
        $\langle \Psi_{6} | \hat A_{2M}^\dag \hat A_{2M} | \Psi_{6} \rangle
            =
        \langle \Psi_{6} | \hat A_{4M}^\dag \hat A_{4M} | \Psi_{6} \rangle $
        $=48/35 \approx 1.37$.

The numerical results for the angular-averaged MF state is shown in Fig \ref{fig2}.
The values oscillates with $\theta$ with period
$\pi/3$ due to the $\cos6\theta$ factor in Eq. (\ref{polar_6}).  For general $\theta$,
the difference between the angular-averaged MF and the spin-singlet pair states is less than $10\%$.\ For example $\langle \hat{A}_{00}^\dagger \hat{A}_{00}\rangle$ $=$ $10.8$ for the exact many-body state while $\langle \hat{A}_{00}^\dagger \hat{A}_{00}\rangle$ $=$ $9.7$ in Fig. \ref{fig2}(a) at $\theta$ $=$ $0$.  The values are identical at $\cos (6 \theta) = -1/3$.

While the density matrices at finite $N$ in general differ, it can be shown \cite{SM} that they have the same leading terms
in the large $N$ limit, so that the energy per particle remains the same up to corrections of order $1/N$, as in the case for spin-1 \cite{HoYip2000}.\ In the large N limit, the fragmented state has macroscopic number fluctuations while they decay rapidly as miniscule magnetization sets in, therefore it is fragile against symmetry-breaking perturbations.\ However we expect that the fragmentation of many-body ground state can be observable in the few-particle system where its signature of two-particle correlations is more noticeable in contrast to the mean-field results.

In conclusion, the many-body ground states of spin-2 Bose gas in zero magnetic field are in general fragmented which is however not describable via angular-averaged MF states.\ For polar states the angular-averaged calculation fails to describe the exact eigenstates when even or odd $N$ $\geq$ $6$ or $9$.\ For cyclic states, the angular-averaged treatment only sustains the exact ground states for particle number of a multiple of $3$, which preserves the constraint of $\langle\hat{\Theta}_2^\dagger\hat{\Theta}_2\rangle$ $=$ $0$.\ That the angular-averaged MF states for even N in spin-1 Bose gas are equivalent to the exact ground states is simply a coincidence.
 For even higher spinor BEC ($f$ $\geq$ $3$), we expect angular-averaged states fail to retrieve the exact eigenstates at
 even smaller number of particles.

{\it Acknowledgments.--} We thank Ryan Barnett and the referee for pointing out to us
\cite{Barnett2011}. This work is supported by the Ministry of Science and Technology, Taiwan,
under grant number MOST-101-2112-M-001-021-MY3.


\clearpage
\section*{Supplemental Material of Fragmented Many-body states of Spin-2 Bose Gas}

In this section, we reproduce the spin-2 irreducible representation of the rotation operator $\hat{D}^{(2)}_{m',m}(\alpha,\beta,\gamma)$ $=$ $e^{-i(m'\alpha+m\gamma)}d^{(2)}_{m',m}(\beta)$ \cite{Brink1968}
for Euler angles $\alpha, \beta, \gamma$. In matrix form, $d^{(2)}_{m',m}(\beta)$ is
\begin{widetext}
\begin{align}
&d^{(2)}_{m',m}(\beta)\nonumber\\&=\left( \begin{array}{ccccc}
\cos^4(\frac{\beta}{2})& -\frac{\sin\beta}{2}(1+\cos\beta)& \sqrt{\frac{3}{8}}\sin^2\beta& \frac{\sin\beta}{2}(\cos\beta-1)& \sin^4\frac{\beta}{2}\\
\frac{\sin\beta}{2}(1+\cos\beta)& \frac{1}{2}(2\cos\beta-1)(\cos\beta+1)& -\sqrt{\frac{3}{2}}\sin\beta\cos\beta& \frac{1}{2}(2\cos\beta+1)(1-\cos\beta)&
\frac{\sin\beta}{2}(\cos\beta-1)\\
\sqrt{\frac{3}{8}}\sin^2\beta& \sqrt{\frac{3}{2}}\sin\beta\cos\beta& \frac{1}{2}(3\cos^2\beta-1)& -\sqrt{\frac{3}{2}}\sin\beta\cos\beta& \sqrt{\frac{3}{8}}\sin^2\beta\\
\frac{\sin\beta}{2}(1-\cos\beta)& \frac{1}{2}(2\cos\beta+1)(1-\cos\beta)& \sqrt{\frac{3}{2}}\sin\beta\cos\beta& \frac{1}{2}(2\cos\beta-1)(\cos\beta+1)& -\frac{\sin\beta}{2}(1+\cos\beta)\\
\sin^4\frac{\beta}{2}& \frac{\sin\beta}{2}(1-\cos\beta)& \sqrt{\frac{3}{8}}\sin^2\beta& \frac{\sin\beta}{2}(\cos\beta+1)& \cos^4(\frac{\beta}{2})\end{array} \right),
\end{align}
\end{widetext}
which is expressed in terms of spin bases $(\varphi_2,$ $\varphi_1,$ $\varphi_0,$ $\varphi_{-1},$ $\varphi_{-2})$.

\section{Angular average mean-field polar states for finite number of particles}
From Eq. (4) in the paper and with
$\hat a^\dag(\hat \Omega)$ $=$ $\sum a^\dag_m \varphi^P_m (\hat \Omega)$ where
$\varphi^P_m(\hat{\Omega})$ $\equiv$ $ {\rm cos} \theta D^{(2)}_{m,0}(\hat{\Omega})$
$+$ $\sin \theta (D^{(2)}_{m,2}(\hat{\Omega})+D^{(2)}_{m,-2}(\hat{\Omega}))/\sqrt{2}$,
we first average over $\alpha$ and $\gamma$, which gives
\begin{align}
&|\Psi'_P(\theta,\beta)\rangle_{av}\nonumber\\
&=\sum_{n=0}^N\binom{N}{n}\left(\sum_{m_1'}\cos\theta d^{(2)}_{m_1',0}(\beta)\hat{a}_{m_1'}^\dagger\right)^n\left(\frac{\sin\theta}{\sqrt{2}}\right)^{N-n}\nonumber\\
&\times \left(\sum_{m_2'}d^{(2)}_{m_2',2}(\beta)\hat{a}_{m_2'}^\dagger+\sum_{m_3'}d^{(2)}_{m_3',-2}(\beta)\hat{a}_{m_3'}^\dagger\right)^{N-n}\nonumber\\
&\times \delta_{f(m'),0} \delta_{f(m),0},
\end{align}
where $f(m)\equiv \sum_{m=m_{1,2,3}}m$.\ We may expand the above further and use one of the delta function constraint $\delta_{f(m),0}$, and the wavefunction becomes
\begin{align}
&|\Psi'_P(\theta,\beta)\rangle_{av}\nonumber\\
&=\sum_{n=0}^N\binom{N}{n}\frac{\cos^n\theta\sin^{N-n}\theta}{2^{(N-n)/2}}\binom{N-n}{\frac{N-n}{2}}\nonumber\\
&\times \left(\sum_{m_1'}d^{(2)}_{m_1',0}(\beta)\hat{a}_{m_1'}^\dagger\right)^n \left(\sum_{m_2'}d^{(2)}_{m_2',2}(\beta)\hat{a}_{m_2'}^\dagger\right)^{\frac{N-n}{2}} \nonumber\\
&\times \left(\sum_{m_3'}d^{(2)}_{m_3',-2}(\beta)\hat{a}_{m_3'}^\dagger\right)^{\frac{N-n}{2}}\delta_{f(m'),0},\label{many2}
\end{align}
where $(N-n)/2$ is integer.\ We then evaluate the $\beta$ average either analytically or with
the help of Mathematica.
In the below, we report the results for this angular-averaged polar states for finite number of particles $N$ $=$ $2$ to $10$.
\subsection{N=2}
From Eq. (\ref{many2}), we have the angular-averaged polar state
\begin{align}
\left|\Psi_P(\theta)\right\rangle_{av}&=\frac{1}{2\sqrt{2!}}\int_0^\pi \left|\Psi'_p(\theta,\beta)\right\rangle \sin\beta d\beta,\nonumber\\
&=\frac{1}{5\sqrt{2}}\left[2\hat{a}_2^\dagger\hat{a}_{-2}^\dagger-2\hat{a}_1^\dagger\hat{a}_{-1}^\dagger+(\hat{a}_0^\dagger)^2\right]|0\rangle,\nonumber\\
&=\frac{1}{5\sqrt{2}}\hat{\Theta}_2^\dagger|0\rangle,
\end{align}
Note that it has no $\theta$ dependence.\ The angular averaged MF state reproduces the exact many-body state.
\subsection{N=3}
From Eq. (\ref{many2}), we have the angular averaged polar state
\begin{align}
\left|\Psi_P(\theta)\right\rangle_{av}&=\frac{1}{2\sqrt{3!}}\int_0^\pi \left|\Psi'_p(\theta,\beta)\right\rangle \sin\beta d\beta,\nonumber\\
&=\frac{2}{35}\left(\cos^3\theta-3\cos\theta\sin^2\theta\right)\hat{\Theta}_3^\dagger|0\rangle\nonumber\\
&=\frac{2}{35}\left[\cos(3\theta)\right]\hat{\Theta}_3^\dagger|0\rangle,
\end{align}
where $\hat{\Theta}_3$ is a three-particle singlet operator,
\begin{align}
\hat{\Theta}_3^\dagger&\equiv \frac{1}{\sqrt{6}}(\hat{a}_0^\dagger)^3-\frac{3}{\sqrt{6}}\hat{a}_1^\dagger\hat{a}_0^\dagger\hat{a}_{-1}^\dagger
+\frac{3}{2}(\hat{a}_1^\dagger)^2\hat{a}_{-2}^\dagger +\frac{3}{2}\hat{a}_{2}^\dagger(\hat{a}_{-1}^\dagger)^2\nonumber\\
&-\frac{6}{\sqrt{6}}\hat{a}_2^\dagger\hat{a}_0^\dagger\hat{a}_{-2}^\dagger.
\end{align}
Note that this angular averaged state has $\theta$ dependence with a period of $2\pi/3$ but always
reproduces the exact many-body state for the region $\gamma < 0$ and near the $+ \beta$ axis.
\subsection{N=4}
For this even number of particles, we again have
\begin{align}
\left|\Psi_P(\theta)\right\rangle_{av}&=\frac{1}{2\sqrt{4!}}\int_0^\pi \left|\Psi'_p(\theta,\beta)\right\rangle \sin\beta d\beta,\nonumber\\
&=\frac{3}{35\sqrt{4!}}\left(\hat{\Theta}_2^\dagger\right)^2|0\rangle,
\label{Psi4av}
\end{align}
where so far we still have a $\theta$ independent angular average.
 This state which is a $N/2$ spin-singlet-pairs state, i.e, the exact eigenstate.
\subsection{N=5}
From Eq. (\ref{many2}), only $n$$=$$1,3,5$ are possible for $(N-n)/2$ is an integer.\ The angular averaged polar state is
\begin{align}
\left|\Psi_P(\theta)\right\rangle_{av}&=\frac{1}{2\sqrt{5!}}\int_0^\pi \left|\Psi'_p(\theta,\beta)\right\rangle \sin\beta d\beta,\nonumber\\
&=\frac{4\sqrt{6}}{77\sqrt{5!}}(\cos^5\theta-3\cos\theta\sin^4\theta\nonumber\\&
-2\cos^3\theta\sin^2\theta)\hat{\Theta}_2^\dagger\hat{\Theta}_3^\dagger|0\rangle\nonumber\\
&=\frac{2}{77\sqrt{5}}\left[\cos(3\theta)\right]\hat{\Theta}_2^\dagger\hat{\Theta}_3^\dagger|0\rangle.
\end{align}
Note that this angular averaged state again has $\theta$ dependence with a period of $2\pi/3$ and
reproduces the exact many-body state for the region $\gamma < 0$ and near the $+ \beta$ axis.
\subsection{N=6}
For this even number of particles, we expect a combination of two- and three-particle singlet states to appear.\ From Eq. (\ref{many2}), we have
\begin{align}
\left|\Psi_P(\theta)\right\rangle_{av}&=\frac{1}{2\sqrt{6!}}\int_0^\pi \left|\Psi'_p(\theta,\beta)\right\rangle \sin\beta d\beta,\nonumber\\
&=\frac{1}{1001\sqrt{6!}}\bigg[(47-2\cos 6\theta)\left(\hat{\Theta}_2^\dagger\right)^3\nonumber\\&
+(12+36\cos 6\theta)\left(\hat{\Theta}_3^\dagger\right)^2\bigg]|0\rangle.
\end{align}

We may express this wavefunction in terms of
the normalized many-body state of two- and three-particle singlet states for N particles,
\begin{align}
\left|\Psi_{N=6}^{(2)} \right\rangle&=\frac{1}{\sqrt{2^4\cdot 3^3\cdot 5\cdot 7}}\left(\hat{\Theta}_2^\dagger\right)^3|0\rangle,\\
\left|\Psi_{N=6}^{(3)} \right\rangle&=\frac{1}{\sqrt{5\cdot 7^3}}\left(\hat{\Theta}_3^\dagger\right)^2|0\rangle,
\end{align}
For $\theta$ $=$ $0$, the normalized angular-averaged state is
\begin{align}
&\left|\tilde{\Psi}_P(0)\right\rangle_{av}\nonumber\\
&=\frac{1}{\sqrt{11\cdot 13\cdot 53}}\left[ 5\cdot 3^{5/2}\left|\Psi_{N=6}^{(2)} \right\rangle +2^2\cdot 7 \left|\Psi_{N=6}^{(3)} \right\rangle\right],
\end{align}
where we note the finite overlap $\langle\Psi_{N=6}^{(2)} |\Psi_{N=6}^{(3)} \rangle$ $=$ $2/(\sqrt{3}\cdot 7)$.\
The angular average has $\theta$ dependence in general, and it can be expressed in terms of two-particle-singlets state only when $\cos6\theta$ $=$ $-1/3$.\ In general the angular-averaged polar state fails to construct the exact ground states which should be $N/2$ spin-singlet-pairs state.
\subsection{N=8}
To investigate the $\theta$ dependence of even number of particles, we proceed to calculate the angular averaged polar state of $N$$=$$8$,
\begin{align}
\left|\Psi_p(\theta)\right\rangle_{av}&=\frac{1}{2\sqrt{8!}}\int_0^\pi \left|\Psi'_p(\theta,\beta)\right\rangle \sin\beta d\beta,\nonumber\\
&=\frac{1}{2431\sqrt{8!}}\bigg[(71-8\cos 6\theta)\left(\hat{\Theta}_2^\dagger\right)^4\nonumber\\&
+16(3+9\cos 6\theta)\hat{\Theta}_2^\dagger\left(\hat{\Theta}_3^\dagger\right)^2\bigg]|0\rangle.
\end{align}

Using the normalized singlet states,
\begin{align}
\left|\Psi_{N=8}^{(2)} \right\rangle&=\frac{1}{\sqrt{2^7\cdot 3^3\cdot 5\cdot 7\cdot 11}}\left(\hat{\Theta}_2^\dagger\right)^4|0\rangle,\\
\left|\Psi_{N=8}^{(2,3)} \right\rangle&=\frac{1}{\sqrt{2\cdot 5\cdot 7\cdot 11\cdot 79}}\hat{\Theta}_2^\dagger\left(\hat{\Theta}_3^\dagger\right)^2|0\rangle,
\end{align}
we may express the normalized angular-averaged state (consider $\theta$ $=$ $0$) as
\begin{align}
\left|\tilde{\Psi}_p(0)\right\rangle_{av}&=\frac{1}{3\sqrt{2\cdot 5}}\left[\sqrt{3}\left|\Psi_{N=8}^{(2)} \right\rangle+\sqrt{79} \left|\Psi_{N=8}^{(2,3)} \right\rangle\right],
\end{align}
where again we use $\langle\Psi_{N=8}^{(2)} |\Psi_{N=8}^{(2,3)} \rangle$ $=$ $4/(\sqrt{3\cdot 79})$.\ Note that when $\cos6\theta$ $=$ $-1/3$, the angular averaged MF state becomes
the exact many-body state.
\subsection{N=10}
We may further investigate the angular averaged MF state for even N.\ From Eq. (\ref{many2}), we have
\begin{align}
\left|\Psi_P(\theta)\right\rangle_{av}&=\frac{1}{2\sqrt{10!}}\int_0^\pi \left|\Psi'_p(\theta,\beta)\right\rangle \sin\beta d\beta,\nonumber\\
&\propto\bigg[(101-20\cos 6\theta)\left(\hat{\Theta}_2^\dagger\right)^5\nonumber\\
&+120(1+3\cos 6\theta)\left(\hat{\Theta}_2^\dagger\hat{\Theta}_3^\dagger\right)^2\bigg]|0\rangle.
\end{align}
Note that a special angle of $\cos6\theta$$=$$-1/3$ appears similar to the cases of $N$$=$$6,8$.

The $\theta$ dependence obtained above can be understood as follows.  In the Cartesian
representation, the general polar state is
$ \cos \theta ( 2 z^2 - x^2 - y^2 ) /\sqrt{6} + \sin \theta (x^2 - y^2)/\sqrt{2}$.
$\theta \to - \theta$ is equivalent to interchanging $x$ and $y$,
whereas $\theta \to  \pi/3 - \theta $ has the effect of interchanging $y$ and $z$
as well as a sign change in the wavefunction.    It follows that the angular
averaged state must be invariant under $\theta \to -\theta$, while
under $\theta \to \pi/3 - \theta$, it is multiplied by $(-1)^N$.
On the other hand, for $N$ particles, the $\theta$ dependence comes
from terms of the form $ \cos^n \theta \sin^{N-n} \theta$ where $n = 0, ...N$ with real coefficients
(see Eq. (\ref{many2})).  Hence it must be of the form
$\sum_{k=-N}^N c_k e^{i k \theta}$ where $c_{-k}=c_k^*$.
For even $N$, it follows that there is no $\theta$ dependence for $N \le 4$,
and the $\theta$ dependence for $6 \le N \le 10$ can only be a linear
combination of a constant and another term $\propto \cos (6 \theta)$
(only $c_{\pm 6}$ and $c_0$ are allowed).  For odd $N$ with $3 \le N \le 7$,
the $\theta$ dependence is via $\cos ( 3 \theta)$ (only $c_{\pm 3}$ allowed).
\section{Angular averaged mean-field cyclic states for finite N}
When we angular averaged the mean-field cyclic state, we obtain, for $N=3$,
\begin{align}
|\Psi_C \rangle_{av} = \frac{4 \sqrt{3}}{ 35 \cdot \sqrt{3!}} \hat \Theta^\dag_3 | 0 \rangle
\end{align}
For  $N$ $=$ $6$, we have
\begin{align}
|\Psi_{C}\rangle_{av} =\frac{8}{7\cdot 11\cdot 13\sqrt{6!}}\left[-\left(\hat{\Theta}_2^\dagger\right)^3+18\left(\hat{\Theta_3^\dagger}\right)^2\right]|0\rangle,
\end{align}
In both cases, we produce the exact many-body states in the C region.
\section{Wavefunctions for the singlet many-body state with $\Lambda=0$ in the C region}
We show how to obtain the singlet wavefunctions in the C regions of Fig 1 in the main text.\ To simplify notations, we shall often simply write $\langle00|n, n'\rangle$ for the Clebsch-Gordan coefficients $\langle00|2n;2n'\rangle$ when no confusion arises.\ We observe that
\begin{align}
\hat\Theta_3^\dag &= c \sum_{m=-2}^2 (-1)^m\hat a_{-m}^\dag\hat A_{2m}^\dag, \nonumber \\
&= \sqrt{5} c \sum_{m=-2}^2 \langle 0 0 | -m, m \rangle \hat a_{-m}^\dag\hat A_{2m}^\dag,\label{T3aA}
\end{align}
where $c = - \frac{1}{2} \left( \frac{7}{3} \right)^{1/2}$.\ Such a relation is expected since both sides create a singlet state of three particles. It is easy to see that
\begin{align}
[\hat a_m, \hat A^{\dag}_{2M} ] &= 2 \langle 2 M | m, M-m \rangle\hat a^{\dag}_{M-m} \nonumber \\
&=  2 (-1)^m \langle 2 \ M-m | -m, M \rangle\hat a^{\dag}_{M-m} \ . \label{amAM}
\end{align}
It is useful to note that though $ [\hat a_m,\hat A_{2m}^{\dag} ] \ne 0$, we have
\begin{align}
\sum_{m=-2}^2 [\hat a_m,\hat A_{2m}^\dag ] = 0.\label{zero}
\end{align}
This relation is expected since the left hand side is rotationally invariant but its right hand side can only involve one creation operator.\ Indeed, $\sum_{m=-2}^2 [ \hat a_m,\hat A_{2m}^\dag ]$ $=$
$2 \sum_{\mu=-2}^2 (-1)^{\mu} \langle 2 0 | \mu, - \mu \rangle a_0^{\dag}$ but the sum is proportional to
$ \sum_{\mu=-2}^2 \langle 0 0 | \mu, - \mu \rangle \langle 2 0 | \mu, - \mu \rangle$ $=$ $0$ due to the orthogonality between the states $ | 0 0 \rangle $ and $ | 2 0 \rangle$.\ From Eq. (\ref{T3aA}) we can evaluate
\begin{align}
[\hat a_{-m},\hat \Theta_3^\dag] = - (-1)^m \frac{ (3 \times 7)^{1/2}}{2}\hat A^\dag_{2, m},
\end{align}
and hence
\begin{align}
 [\hat\Theta_2,\hat \Theta_3^{\dag} ] &= -  \frac{ (3 \times 7)^{1/2}}{2} \sum_{m=-2}^2 \{\hat A_{2m}^\dag,\hat a_m \}, \nonumber \\
 &= -  (3 \times 7)^{1/2} \sum_{m=-2}^2\hat  A_{2m}^\dag\hat a_m,
\end{align}
where in the last step we have used Eq. (\ref{zero}).

From the above we find
\begin{align}
\hat{\Theta}_2 (\hat{\Theta}_2^{\dag})^Q \hat{\Theta}_3^{\dag R} | 0 \rangle &= 2 Q ( 6 R + 2Q +3) (\hat{\Theta}_2^\dag)^{Q-1} (\hat{\Theta}_3^{\dag})^{R} | 0 \rangle \nonumber\\
&+\frac{3 R (R-1)} {2} (\hat{\Theta}_2^\dag)^{Q+2} (\hat{\Theta}_3^{\dag})^{R-2} | 0 \rangle.
\label{T2QR}
\end{align}
Note that this implies, for the special case $R = 0$, 
\begin{align}
\hat{\Theta}_2^\dag \hat{\Theta}_2 (\hat{\Theta}_2^{\dag})^Q | 0 \rangle = 
2 Q ( 2Q +3) (\hat{\Theta}_2^\dag)^{Q} | 0 \rangle \ ,
\end{align}
a result which we shall see again in Sec \ref{RDM} of this SM.

We can now derive the exact manybody wavefunction for the region C when $N$ is a multiple of $3$.
The state $|\Psi_{3R} \rangle $ with $N=3R$ particles and
 $\tau$ $=$ $N$ with $N$ $=$ $3R$ being a multiple of $3$ ({\it i.e.} $\hat{\Theta}_2^{\dag} \hat{\Theta}_2 | \Psi_{3R} \rangle$ $=$ $0$ hence $\Lambda = 0$) can then be constructed as
\begin{align}
|\Psi_{3R}\rangle&=b_0(\hat{\Theta}_3^\dagger)^R+b_1(\hat{\Theta}_2^\dagger)^3(\hat{\Theta}_3^\dagger)^{R-2}+...\nonumber\\
&+b_k(\hat{\Theta}_2^\dagger)^{3k}(\hat{\Theta}_3^\dagger)^{R-2k}+...,
\end{align}
where we have
\begin{align}
\frac{b_{k+1}}{b_k}=-\frac{(R-2k)(R-2k-1)}{12(k+1)(2R-2k-1)}.
\end{align}

We also note here that since $[\Theta_2, N(N+3) - \Theta_2^{\dag} \Theta_2 ]$ $=$ $0$, the states $(\Theta_2^{\dag})^Q | \Psi_{3R} \rangle$ have the same quantum number $\tau$ $=$ $3R$ though different particle numbers $N$ $=$ $2Q$ $+$ $3R$.\
From these we obtain the exact many-body ground states in region C of Fig. 1  in the main text.
\section{Reduced density matrix calculation}\label{RDM}
We here consider the $N=2Q$ singlet state
\begin{align}
| \Psi_{2Q} \rangle = \frac{1}{\sqrt{g(Q)}} \hat{\Theta}_2^{\dag}{}^Q |0\rangle,\label{2QS}
\end{align}
where $g(Q)$ is a normalization constant.\ $g(Q)$ can be evaluated (see also below) by the repeated use of the commutation relation $[\hat{\Theta},\hat{\Theta}^\dag] = 2 ( 2 \hat N + 5)$ where $\hat N$ is the number operator.\ We then obtain
\begin{align}
g(Q) = 2^Q Q! (2Q + 3)!! / 3 \label{gQ}.
\end{align}
Some special values are: $g(1)$ $=$ $10$, $g(2)$ $=$ $2^3 \times 5 \times 7$, and $g(3)$ $=$ $2^4 \times 3^3 \times 5 \times 7$.\ It turns out that $g(Q)$ $=$ $f(1, Q)$ of \cite{HoYip2000}.

     The expectation values
     for the two-particle density matrices
     $ \langle \Psi_{2Q} | \hat A_{JM}^\dag \hat A_{JM} | \Psi_{2Q} \rangle $
      needed can be read off from the energy
      $E = \frac{1}{2} \left[ \alpha N (N-1) + \beta ( F (F+1) - 6 N) + \gamma \Lambda \right]$ where $\Lambda = 2Q ( 2Q+3)$
      since this must also be
       $E = \frac{1}{2} \sum_{JM} g_F \langle \Psi_{2Q} | \hat A_{JM}^\dag \hat A_{JM} | \Psi_{2Q} \rangle$
      where the sum over $J$ is for $0,2,4$ only.  We have
      \begin{align}
        \langle \Psi_{2Q} | \hat A_{00}^\dag \hat A_{00} | \Psi_{2Q} \rangle
         &= 2Q (2Q +3) / 5  \label{DM0}  \\
        \langle \Psi_{2Q} | \hat A_{2M}^\dag \hat A_{2M} | \Psi_{2Q} \rangle
         &= 8Q (Q-1) / 35
         \label{DM2}  \\
        \langle \Psi_{2Q} | \hat A_{4M}^\dag \hat A_{4M} | \Psi_{2Q} \rangle
         &= 8Q (Q-1) / 35
         \label{DM4}
      \end{align}
     The equality between the values
      between $J=2$ and $J=4$ is due to the special properties
      of the state $ | \Psi_{2Q} \rangle $. In below we also show an alternate derivation of Eq. (\ref{DM0}-\ref{DM4})

The state (\ref{2QS}) can be expressed in terms of the basis $ | n_2, n_1, n_0, n_{-1}, n_{-2} \rangle $, where $n_{m}$ is the number of particles in the state $m$.\ We get
\begin{align}
| \Psi_{2Q} \rangle&= \left[ \frac{ 3 \times 2^Q Q! } { (2Q+3)!!} \right]^{1/2}\sum_{k_0, k_1, k_2}^Q{}'(-1) ^{k_1} \frac { [ ( 2 k_0 )! ] ^{1/2} }{ 2^{k_0} k_0! }
\nonumber\\
&\times | k_2, k_1, 2k_0, k_1, k_2 \rangle \label{2QS2},
\end{align}
where the sum is over all non-negative integers $k_0$, $k_1$, $k_2$ with the restriction (denoted by the prime) $k_0+ k_1 + k_2 = Q$. \ The density matrices are obtained by operating
$\hat{a}_{m_1} \hat{a}_{m_2} $ on $ | \Psi_{2Q} \rangle$ and then evaluating the appropriate inner products.\ The required sums are evaluated below.\

We show here how to evaluate the sums involved.\ They are of the form
\begin{align}
S_Q \equiv \sum_{k=0}^Q b_k,
\end{align}
where $b_k$ are the products of polynomials in $k$ with $c_k \equiv \frac { (  2 k )! }{ 2^{2k} (k!)^2} = \frac{ (2k-1)!!} { 2^k k! } $.\ For this, we notice that if the function $f(y) \equiv \sum_{k=0}^\infty b_k y^k$ is known, then (by straight-forward verification) $S_Q$ is simply the coefficient of $y^Q$ of the function $F(y) \equiv f(y)/ ( 1 - y)$.\ Now, we note that $f_1(y) \equiv  \sum_{k=0}^\infty c_k y^k$ is given simply by $ ( 1 - y)^{-1/2}$.\ Hence the sum
$S_{1,Q} \equiv \sum_{k=0}^Q c_k$ is given by the $y^Q$ coefficient of $ (1 - y)^{-3/2}$, and hence
\begin{align}
S_{1,Q} \equiv \sum_{k=0}^Q c_k = \frac{ ( 2Q + 1)!!} { 2^Q Q! }.
\end{align}
Similarly, for the sum  $S_{2,Q} \equiv \sum_{k=0}^Q k c_k$, the function $f_2(y)  \equiv \sum_{k=0}^\infty k c_k y^k$ can be obtained from $ y \frac{d}{dy} f_1(y) = \frac{y}{2} (1 - y)^{-3/2}$.\ Hence $S_{2,Q}$ is the $y^Q$ coefficient of $F_{2}(y) = \frac{y}{2} ( 1 - y)^{-5/2}$, and hence
\begin{align}
S_{2,Q} \equiv \sum_{k=0}^Q k c_k =  \frac{Q}{3} \frac{ ( 2Q + 1)!!} { 2^Q Q! }.
\end{align}
We can proceed similarly to get
\begin{align}
S_{3,Q} &\equiv \sum_{k=0}^Q k (k-1)  c_k =  \frac{Q(Q-1)}{5} \frac{ ( 2Q + 1)!!} { 2^Q Q! },\\
S_{4,Q} &\equiv \sum_{k=0}^Q k (k-1) (k-2) c_k, \nonumber\\
&= \frac{Q(Q-1)(Q-2)}{7} \frac{ ( 2Q + 1)!!} { 2^Q Q! }.
\end{align}
With the above sums, we can also obtain
\begin{align}
\sum_{k=0}^Q  (Q + 1 - k)  c_k &=  \frac{1}{3} \frac{ ( 2Q + 3)!!} { 2^Q Q! },\\ \label{SQ1k}
\sum_{k=0}^Q (Q - k ) (Q - k + 1)  c_k &=  \frac{4 Q}{15} \frac{ ( 2Q + 3)!!} { 2^Q Q! },
\end{align}
(using $ ( Q - k) ( Q - k + 1)$ $=$ $k (k-1 )$ $-$ $2Qk$ $+$ $Q(Q+1)$), and
\begin{align}
&\sum_{k=0}^Q (Q-k-1) (Q - k ) (Q - k + 1)  c_k \nonumber\\
&=\frac{ 8Q (Q-1)}{35} \frac{ ( 2Q + 3)!!} { 2^Q Q! },
\end{align}
(using $ (Q - k - 1)( Q - k) ( Q - k + 1) = - k(k-1)(k-2) + 3 (Q-1) k (k-1 ) - 3Q(Q-1)k + (Q-1)Q(Q+1)$).

We demonstrate the use of the above relations by checking here the normalization of $| \Psi_{2Q} \rangle $.\ $ \langle \Psi_{2Q} | \Psi_{2Q} \rangle $ is given by $ \left[ \frac{ 3 \times 2^Q Q! } { (2Q+3)!!} \right] \sum_{k_0, k_1, k_2}^Q {}' c_{k_0} $.\ Due to the restriction $k_0+k_1+k_2 = Q$, the sum is therefore given by $\sum_{k_0=0}^Q [\sum_{k_1 = 0}^{Q-k_0} 1 ]$ $= \sum_k^Q c_k (Q-k+1)$, which is $ \frac  { (2Q+3)!!}{ 3 \times 2^Q Q! }$ from Eq. (\ref{SQ1k}).\ The density matrices are obtained by first operating $\hat a_m$ or $\hat a_{m_1}\hat a_{m_2} $ on $ | \Psi_{2Q} \rangle$ and then evaluating the appropriate inter-products with the help of the above formulas.

We list here also the two-particle density matrices $ \langle \Psi_{2Q} |\hat a_{m_1}^\dag\hat a_{m_2}^\dag\hat a_{m_3}\hat a_{m_4} | \Psi_{2Q} \rangle $.\ We list them starting from the largest $M$ $\equiv$ $m_1$ $+$ $m_2$ $=$ $m_3$ $+$ $m_4$.\ For M = 4,
\begin{align}
\langle \Psi_{2Q} |\hat a_{2}^\dag\hat a_{2}^\dag\hat a_{2}\hat a_{2} | \Psi_{2Q} \rangle = 8 Q (Q-1)/35.
\end{align}
For M = 3
\begin{align}
\langle \Psi_{2Q} |\hat a_{2}^\dag \hat a_{1}^\dag\hat a_{1}\hat a_{2} | \Psi_{2Q} \rangle= 4 Q (Q-1)/35,
\end{align}
corresponding to $\langle \Psi_{2Q} |\hat A_{4M}^\dag\hat A_{4M} | \Psi_{2Q} \rangle = 8 Q (Q-1)/35$.\ For M=2, we have two operators $\hat a_0 \hat a_2$ and $\hat a_1 \hat a_1$ and their conjugates.\ We obtain
\begin{align}
\langle \Psi_{2Q} |\hat a_{2}^\dag\hat a_{0}^\dag\hat a_{0}\hat a_{2} | \Psi_{2Q} \rangle&= 4 Q (Q-1)/35,\\
\langle \Psi_{2Q} |\hat a_{1}^\dag\hat a_{1}^\dag\hat a_{1}\hat a_{1} | \Psi_{2Q} \rangle&= 8 Q (Q-1)/35,
\end{align}
whereas
\begin{align}
\langle \Psi_{2Q} |\hat a_{2}^\dag\hat a_{0}^\dag\hat a_{1}\hat a_{1} | \Psi_{2Q} \rangle= 0.
\end{align}
The last result can be most easily seen when $ \Psi_{2Q} $ is expanded in the number basis.\ This is reflected in the equality between Eqs. (\ref{DM2}) and (\ref{DM4}).

For M=1, there are two operators $\hat a_1\hat a_0$ and $\hat a_2\hat a_{-1}$.  We have
\begin{align}
\langle \Psi_{2Q} |\hat a_{1}^\dag\hat a_{0}^\dag\hat a_{0}\hat a_{1} | \Psi_{2Q} \rangle
&=   \langle \Psi_{2Q} |\hat a_{2}^\dag\hat a_{-1}^\dag\hat a_{-1}\hat a_{2} | \Psi_{2Q} \rangle, \\
&= 4 Q (Q-1)/35, \end{align}
and there are no cross elements.\ Similar remarks we made for the $M=2$ sector also applies here.

For M=0, there are three operators, $\hat a_2\hat a_{-2}$ and $\hat a_1\hat a_{-1}$ and $\hat a_0\hat a_0$.\ This part of the density matrix is, with rows and columns in order of these three operators, given by
\begin{align}
\frac{2Q}{35}
\left( \ba{ccc}
(4Q+3) & -(2Q+5) &  (2Q+5) \\
-(2Q+5) &  (4Q+3) & -(2Q+5) \\
(2Q+5)  &   -(2Q+5) & (6Q+1)
\ea\right).
\end{align}
The equality between the first two diagonal elements, as well as among the off-diagonal elements except signs, follows from the fact that $| \Psi_{2Q} \rangle$ is invariant under $\hat a_{\pm 2} \to\hat a_{\pm 1}$ up to a sign.
From the above formulas, we can recover Eqs. (\ref{DM0})-(\ref{DM4}).

It is straightforward to obtain the number fluctuations from above.\ $ \langle \hat N_2 \hat N_2 \rangle = \langle \hat N_1 \hat N_1 \rangle = \langle \hat N_2 \hat N_{-2} \rangle = \langle \hat N_1 \hat N_{-1} \rangle = 2Q (4Q + 3) / 35$, $ \langle \hat N_0 \hat N_0  \rangle = 4Q (3Q + 4) / 35$,
$ \langle \hat N_2 \hat N_1 \rangle = \langle \hat N_2 \hat N_0 \rangle = \langle \hat N_2 \hat N_{-1} \rangle = \langle \hat N_1 \hat N_0 \rangle$ $= 4Q (Q-1)/35$.\ The above expressions are valid also when replacing $m$ by $-m$.\ $ \langle \hat N_2 \hat N_2 \rangle =  \langle \hat N_2 \hat N_{-2} \rangle $ etc follows immediately also from (\ref{2QS2}), since for each of the states on the right-hand-side, $n_2 = n_{-2}$.

We now consider the density matrices for $|\Psi_P\rangle_{av}$ in large $N$ limit.\ We have
\begin{widetext}
\begin{align}
\langle \hat{a}^\dag_{m_1}\hat{a}^\dag_{m_1} \hat{a}_{m_3} \hat{a}_{m_4} \rangle = N (N -1 )
\frac{ \int_{\hat \Omega_1,\hat \Omega_2} \varphi_{m_1}^*(\hat \Omega_1) \varphi_{m_2}^*(\hat \Omega_1)\varphi_{m_3}(\hat \Omega_2) \varphi_{m_4} (\hat \Omega_2)
\left[ \sum_m \varphi_m^*(\hat \Omega_1) \varphi_m (\hat \Omega_2) \right]^{N-2} }
{ \int_{\hat \Omega_1,\hat \Omega_2} \left[ \sum_m \varphi_m^*(\hat \Omega_1) \varphi_m (\hat \Omega_2) \right]^{N} }.
\end{align}
\end{widetext}
(We leave out the explicit labels $|\Psi_P\rangle_{av}$ to simplify our notations).
For large $N$, the overlap $ [\sum_m \varphi_m^* (\hat \Omega_1)\varphi_m (\hat \Omega_2) ] ^N$ is negligible unless $\hat \Omega_1$ is very close to $\hat \Omega_2$.
(rigorously speaking, also $\hat \Omega_2 ~ - \hat \Omega_1$, but we can check easily that this does not
affect the following argument).\ Hence we can identify the $\hat \Omega$'s in the arguments of $\varphi_{m_1}$ $...$ $\varphi_{m_4}$ in the integrand of the numerator.\ Canceling the common factors (the normalization coefficient is $\propto N^{-1}$ for large $N$) we are left with
\begin{align}
&\langle \hat{a}^\dag_{m_1} \hat{a}^\dag_{m_1} \hat{a}_{m_3}\hat a_{m_4} \rangle \nonumber\\
&=N (N -1 )\int_{\hat \Omega} \varphi_{m_1}^*(\hat \Omega) \varphi_{m_2}^*(\hat \Omega) \varphi_{m_3}(\hat \Omega) \varphi_{m_4} (\hat \Omega). \label{DMNinfty}
\end{align}
Now we observe that, for our state, $\varphi_m^*(\hat \Omega)$ $=$ $(-1)^m \varphi_{-m}(\hat \Omega)$, since $d^{(2)}_{-m, -n}$ $=$ $(-1)^{m+n} d^{(2)}_{m, n}$ where $\hat{D}^{(2)}_{m',m}(\alpha,\beta,\gamma)$ $\equiv$ $e^{-i(m'\alpha+m\gamma)}d^{(2)}_{m',m}(\beta)$.\ Therefore the integral above is the same as
\begin{align}
(-1)^{m_1+m_2} \int_{\hat \Omega} \varphi_{-m_1}(\hat \Omega) \varphi_{-m_2}(\hat \Omega) \varphi_{m_3}(\hat \Omega) \varphi_{m_4} (\hat \Omega).\nonumber
\end{align}
We notice that this latter integral is the same as the one that occurs in our evaluation of the coefficient of $\hat{a}^\dag_{-m_1} \hat{a}^\dag_{-m_2} \hat{a}^{\dag}_{m_3} \hat{a}^\dag_{m_4}$ for the wavefunction $| \Psi_P \rangle_{av} $ for four particles, except combinatorial factors.\ For example, the value of $ \langle \hat{a}^\dag_2 \hat{a}^\dag_1 \hat{a}_1 \hat{a}_2 \rangle $ is just $-N(N-1)$ times the coefficient of $ \hat{a}^\dag_{-2} \hat{a}^\dag_{-1} \hat{a}^\dag_1 \hat{a}^\dag_2 $ in $| \Psi_P \rangle_{av} $ (see Eq. (\ref{Psi4av})) divided by $ 4!$.\ Using our previous calculations we therefore obtain, in the large $N$ limit, $ \langle \hat{a}^\dag_2 \hat{a}^\dag_1 \hat{a}_1 \hat{a}_2 \rangle$ $=$ $N^2 / 35$, $ \langle \hat{a}^\dag_2 \hat{a}^\dag_0 \hat{a}_0 \hat{a}_2 \rangle$ $=$ $N^2 / 35$ etc.\ It is again most economical to express the final results using $\hat A_{JM}$.\ We get $\langle\hat A^\dag_{00}\hat A_{00} \rangle$ $=$ $N^2/5$ and $\langle\hat A^\dag_{2M}\hat A_{2M} \rangle$ $=$ $\langle\hat A^\dag_{4M}\hat A_{4M} \rangle$ $=$ $2N^2 / 35$.\ Hence the $N^2$ terms in two-particle density matrix in the state obtained by angular average is the same as that of Eq. (\ref{2QS}), and the differences arise only in lower powers in $N$.\ Therefore the interaction energies per particle for these states are equal except for terms that are of order $1$.

\begin{thebibliography}{99}
\bibitem{Stamper-Kurn1998} D.M. Stamper-Kurn, M.R. Andrews, A.P. Chikkatur, S. Inouye, H.-J. Miesner, J. Stenger, W. Ketterle, Phys. Rev. Lett. {\bf 80}, 2027 (1998).
\bibitem{Stenger1998} J. Stenger, S. Inouye, D.M. Stamper-Kurn, H.-J. Miesner, A.P. Chikkatur, W. Ketterle, Nature {\bf 396}, 345 (1998).
\bibitem{Kawaguchi2012} Y. Kawaguchi and M. Ueda, Phys. Report {\bf 520}, 253 (2012).
\bibitem{Stamper-Kurn2013} D.M. Stamper-Kurn and M. Ueda, Rev. Mod. Phys. {\bf 85}, 1191 (2013).
\bibitem{Ho1998} T.-L. Ho, Phys. Rev. Lett. {\bf 81}, 742 (1998).
\bibitem{Ohmi1998} T. Ohmi, K. Machida, J. Phys. Soc. Japan {\bf 67}, 1822 (1998).
\bibitem{Ciobanu2000} C.V. Ciobanu, S.-K. Yip, T.-L. Ho, Phys. Rev. A {\bf 61}, 033607 (2000).
\bibitem{Koashi2000} M. Koashi and M. Ueda, Phys. Rev. Lett. {\bf 84}, 1066 (2000).
\bibitem{Ueda2002} M. Ueda, M. Koashi, Phys. Rev. A {\bf 65}, 063602 (2002).
\bibitem{Turner2007} A. M. Turner, R. Barnett, E. Demler, and A. Vishwanath, Phys. Rev. Lett. {\bf 98}, 190404 (2007).
\bibitem{Song2007} J. L. Song, G. W. Semenoff, and F. Zhou, Phys. Rev. Lett. {\bf 98}, 160408 (2007).
\bibitem{Diener2006} R.B. Diener and T.-L. Ho, Phys. Rev. Lett. {\bf 96}, 190405 (2006).
\bibitem{Santos2006} L. Santos and T. Pfau, Phys. Rev. Lett. {\bf 96}, 190404 (2006).
\bibitem{Yip2007} S.-K. Yip, Phys. Rev. A {\bf 75}, 023625 (2007).
\bibitem{Kawaguchi2011} Y. Kawaguchi, M. Ueda, Phys. Rev. A {\bf 84}, 053616 (2011).
\bibitem{Chang2004} M.-S. Chang, C.D. Hamley, M.D. Barrett, J.A. Sauer, K.M. Fortier, W. Zhang, L. You, M.S. Chapman, Phys. Rev. Lett. {\bf 92}, 140403 (2004).
\bibitem{Black2007} A.T. Black, E. Gomez, L.D. Turner, S. Jung, P.D. Lett, Phys. Rev. Lett. {\bf 99}, 070403 (2007).
\bibitem{Schmaljohann2004} H. Schmaljohann, M. Erhard, J. Kronjager, M. Kottke, S. van Staa, L. Cacciapuoti, J.J. Arlt, K. Bongs, K. Sengstock, Phys. Rev. Lett. {\bf 92}, 040402 (2004).
\bibitem{Kuwamoto2004} T. Kuwamoto, K. Araki, T. Eno, T. Hirano, Phys. Rev. A {\bf 69}, 063604 (2004).
\bibitem{Guzman2011} J. Guzman, G.-B. Jo, A. N. Wenz, K. W. Murch, C. K. Thomas, and D. M. Stamper-Kurn, Phys. Rev. A {\bf 84}, 063625 (2011).
\bibitem{Kronjager2006} J. Kronj\"{a}ger, C. Becker, P. Navez, K. Bongs, and K. Sengstock, Phys. Rev. Lett. {\bf 97}, 110404 (2006).
\bibitem{Klempt2009} C. Klempt, O. Topic, G. Gebreyesus, M. Scherer, T. Henninger, P. Hyllus, W. Ertmer, L. Santos, and J. J. Arlt, Phys. Rev. Lett. {\bf 103}, 195302 (2009).
\bibitem{Liu2009} Y. Liu, S. Jung, S. E. Maxwell, L. D. Turner, E. Tiesinga, and P. D. Lett, Phys. Rev. Lett. {\bf 102}, 125301 (2009).
\bibitem{Jiang2014} J. Jiang, L. Zhao, M. Webb, and Y. Liu, Phys. Rev. A {\bf 90}, 023610 (2014).
\bibitem{Sadgrove2013} M. Sadgrove, Y. Eto, S. Sekine, H. Suzuki, and T. Hirano, J. Phys. Soc. Jpn. {\bf 82}, 094002 (2013).
\bibitem{Eto2014} Y. Eto, M. Sadgrove, S. Hasegawa, H. Saito, and T. Hirano, Phys. Rev. A {\bf 90}, 013626 (2014).
\bibitem{Penrose1956} O.Penrose and L. Onsager, Phys. Rev. {\bf 104}, 576 (1956).
\bibitem{Nozieres1982} P. Nozi\`{e}res and D. Saint James, J. Phys. (Paris) {\bf 43}, 1133 (1982).
\bibitem{HoYip2000} T.-L. Ho and S.-K. Yip, Phys. Rev. Lett. {\bf 84}, 4031 (2000).
\bibitem{Klaiman2014} S. Klaiman, A. U. J. Lode, A. I. Streltsov, L. S. Cederbaum, and O. E. Alon, Phys. Rev. A {\bf 90}, 043620 (2014).
\bibitem{Tasaki2013} H. Tasaki, Phys. Rev. Lett. {\bf 110}, 230402 (2013).
\bibitem{Sarlo2013} L. D. Sarlo, L. Shao, V. Corre, T. Zibold, D. Jacob, J. Dalibard and F. Gerbier, New J. Phys. {\bf 15}, 113039 (2013).
\bibitem{Bader2009} P. Bader and U. R. Fischer, Phys. Rev. Lett. {\bf 103}, 060402 (2009).
\bibitem{Cizek2013} N. C. Cizek and M. A. Kasevich, Phys. Rev. A {\bf 88}, 063641 (2013).
\bibitem{Kang2014} M.-K. Kang and U. R. Fischer, Phys. Rev. Lett. {\bf 113}, 140404 (2014).
\bibitem{Kawaguchi2014} Y. Kawaguchi, Phys. Rev. A {\bf 89}, 033627 (2014).
\bibitem{Ozawa2012-Zhou2013-Song2014} T. Ozawa and G. Baym, Phys. Rev. A {\bf 85}, 013612 (2012); Q. Zhou and X. Cui, Phys. Rev. Lett. {\bf 110}, 140407 (2013); S.-W. Song, Y.-C. Zhang, H. Zhao, X. Wang, and W.-M. Liu, Phys. Rev. A {\bf 89}, 063613 (2014).
\bibitem{Law1998} C. K. Law, H. Pu, and N. P. Bigelow, Phys. Rev. Lett. {\bf 81}, 5257 (1998).
\bibitem{Mueller2006} E. J. Mueller, T.-L. Ho, M. Ueda, and G. Baym, Phys. Rev. A {\bf 74}, 033612 (2006).

\bibitem{Mermin1974} N. D. Mermin, Phys. Rev. B {\bf 9}, 869 (1974).
\bibitem{HoYin2000} T.-L. Ho and L. Yin, Phys. Rev. Lett. {\bf 84}, 2302 (2000).
\bibitem{Uchino2008} S. Uchino, T. Otsuka, and M. Ueda, Phys. Rev. A {\bf 78}, 023609 (2008).
\bibitem{SM} Supplemental Materials.
\bibitem{Brink1968} D.M. Brink and G.R. Satchler, {\it Angular Momentum} (Oxford University Press, 1968).
\bibitem{Barnett2011} R. Barnett, H.-Y. Hui, C.-H. Lin, J. D. Sau, and S. Das Sarma, Phys. Rev. A {\bf 83}, 023613 (2011).
\end{thebibliography}
\end{document}